\documentclass[lettersize,journal]{IEEEtran}
\usepackage{stfloats}
\usepackage{cite}
\usepackage{amsmath,amssymb,amsfonts}
\usepackage{graphicx}
\usepackage{textcomp}
\usepackage{xcolor}
\def\BibTeX{{\rm B\kern-.05em{\sc i\kern-.025em b}\kern-.08em
    T\kern-.1667em\lower.7ex\hbox{E}\kern-.125emX}}
\usepackage{amsmath}   
\usepackage{verbatim}

\usepackage[utf8]{inputenc}

\usepackage{epstopdf}
\usepackage{stmaryrd}
\usepackage{multirow,url}
\usepackage{subfigure}
\usepackage{enumerate}
\usepackage{amsthm}

\begin{document}

\title{Low-Power Timely Random Access: Packet-based or Connection-based?}

\author{Tse-Tin~Chan,~\IEEEmembership{Member,~IEEE,}~Jian~Feng,~Haoyuan~Pan,~\IEEEmembership{Member,~IEEE}

\thanks{This work was supported by the Guangdong Basic and Applied Basic Research Foundation under Grant 2021A1515012601.}
\thanks{T.-T.~Chan is with the Department of Mathematics and Information Technology, The Education University of Hong Kong, Hong Kong SAR, China (e-mail: {tsetinchan@eduhk.hk}).}
\thanks{J. Feng and H. Pan are with the College of Computer Science and Software Engineering, Shenzhen University, Shenzhen, 518060, China (e-mails: {fengjian2020@email.szu.edu.cn}, {hypan@szu.edu.cn}). }
}
\maketitle

\begin{abstract}
This paper studies low-power random access protocols for timely status update systems with information freshness requirements, measured by age of information (AoI). In an extensive network, a fundamental challenge is scheduling a large number of transmitters to access the wireless channel in a way that achieves low network-wide AoI while consuming minimal power. Conventional \emph{packet-based} random access protocols involve transmitters contending for the channel by sending their entire data packets. When the packet duration is long, the time wasted due to packet collisions can be significant. In contrast, \emph{connection-based} random access protocols establish connections with the receiver before transmitting data packets. From an information freshness perspective, there should be conditions that favor one approach over the other. We present a comparative study of the average AoI of packet-based and connection-based random access protocols. Specifically, we consider frame slotted Aloha (FSA) as a representative of packet-based random access and design a request-then-access (RTA) protocol for connection-based random access. Our analyses indicate that the choice between packet-based or connection-based protocols depends mainly on the payload size of update packets and the transmit power budget. In particular, RTA saves power and significantly reduces AoI, especially when the payload size is large. Overall, our investigation offers insights into the practical design of random access protocols for low-power timely status update systems.
\end{abstract}





\section{Introduction}\label{sec:introduction}

The next-generation Internet of Things (IoT) network has been envisioned as a key enabler for emerging applications such as connected vehicles in smart cities and cyber-physical systems in the industry. Empowered by IoT, billions of connected devices with sensing, monitoring, and communication capabilities operate collaboratively and intelligently to enable reliable real-time communication, interactions, and decision-making \cite{AoIMagazine,urllc_popo}. In these applications, providing fresh status updates is of utmost importance. For example, periodic safety message exchange in vehicular networks should be delivered among vehicles as timely as possible to ensure safety \cite{aoi_vehicular}. 

Age of information (AoI) has been a key performance metric for measuring information freshness \cite{aoi_mono,AoI,yates2021age}. It is defined as the time elapsed since the generation time of the latest received information update at the destination. Specifically, suppose that the latest information update received by the receiver is the update packet generated by the source at time $t'$, then the instantaneous AoI at time $t$ is $t-t'$. Previous studies have shown that AoI depends on the data generation pattern and transmission delay through the network \cite{yates2021age}. As a result,  AoI differs significantly from conventional metrics such as delay and latency \cite{aoi_mono,AoI,yates2021age} and has attracted considerable research interest, particularly for IoT applications requiring timely data.

To achieve low network-wide AoI for large-scale wireless IoT systems, a fundamental design challenge is scheduling massive IoT devices to access the wireless channel, especially when these devices are typically power-constrained, such as tiny sensors. Scheduled access protocols generally require centralized coordination and decision-making, which can be restrictive for many IoT scenarios with a large number of low-cost sensors \cite{aoi_tdma_fdma}. Hence, efficient random access protocols that operate in a distributed and decentralized manner have been extensively studied. For example, despite its simplicity, \cite{yang2021understanding} demonstrated the effectiveness of slotted Aloha (SA) in achieving low AoI in massive access networks. In random access, packet collisions often occur when multiple transmitters send packets concurrently due to a lack of coordination, resulting in high AoI and power wastage. While multiple packet replicas can be sent so that advanced interference cancellation techniques can recover the original packets and improve AoI performance \cite{aoi_irsa}, transmitting multiple replicas leads to high power consumption.  This paper aims to examine low-power random access protocols for timely status update systems with AoI requirements.

In conventional random access schemes, a transmitter contends for the channel by sending its entire data packet, known as packet-based random access \cite{kurose2007computer,gao2019random}. Typical packet-based random access schemes include SA and the distributed coordination function (DCF) in WiFi networks \cite{dot11std13}. Since no transmitters can successfully update when more than one transmitter sends simultaneously, the time wasted by a failed transmission can be significant when the data packet duration is long. In contrast to packet-based random access, connection-based random access allows a transmitter to send a short request frame to contend for the channel first, rather than sending the data packet directly \cite{gao2019random}. The transmitter can only send the data packet if it successfully receives an acknowledgment from the receiver. The request-to-send/clear-to-send (RTS/CTS) access mechanism in WiFi networks is a typical example of connection-based random access: a transmitter sends an RTS frame and waits for a CTS frame from the receiver to establish a connection before transmitting data packets \cite{dot11std13}. Thanks to the established connection, the data packet transmission is usually collision-free. Consequently, transmission failures involve only request frames and the wasted time could be much shorter than that of a data packet. In other words, if the data packet is long enough, the transmission failure time and the wasted transmit power can be reduced, i.e., the cost of connection establishment is relatively small.

Developing AoI-aware random access schemes for timely status update systems that consider limited transmit power is crucial. In particular, properly using packet-based or connection-based random access protocols is of great practical importance.  Intuitively, there should be a critical threshold for the data packet duration beyond which it is beneficial to establish a connection, e.g., it should be longer than the request frame to some extent \cite{gao2019random}. Previous works have not investigated the AoI-aware characterization of such a threshold, and this paper aims to fill this research gap. We present a comparative study of the average AoI of packet-based and connection-based random access protocols in the context of an average transmit power budget. 

We first consider frame slotted Aloha (FSA) as a representative of packet-based random access protocols \cite{kurose2007computer}. In FSA, time is divided into frames, each containing a fixed number of time slots. If a transmitter wants to send a packet in a frame, it randomly chooses one of the time slots to transmit. To analyze connection-based random access, we design a request-then-access (RTA) protocol inspired by FSA. RTA consists of two phases. In the first phase (the request phase), transmitters contend and send update request frames to establish a connection with the receiver. In the second phase (the access phase), only transmitters that successfully contended in the request phase send update packets in a TDMA superframe, which is collision-free.

The AoI analysis of RTA is more complicated than that of FSA. For example, the number of transmitters entering the access phase is random and depends on the probability of sending an update request during the request phase. More specifically, when the probability of sending an update request is too high or too low, the number of transmitters entering the access phase will be small due to high collision probability or low request update rate. Additionally, if a large number of transmitters enter the access phase, the duration of the TDMA superframe will also be long. In this case, the time required between two successful updates could also be long, which also affects the average AoI. Therefore, the probability of sending an update request in RTA should be carefully designed to minimize the average AoI. 

We derive the closed-form average AoI and average transmit power consumption formulas of FSA and RTA. Our simulations show that whether to use packet-based or connection-based random access protocols mainly depends on the payload size of update packets and the transmit power budget of transmitters. When the payload size is large, RTA often outperforms FSA because RTA dedicates a connection establishment phase to help avoid collisions of data packets, thus saving power and reducing AoI. When the duration of an update packet and a request frame are comparable (e.g., 16 bytes), RTA should still be used when the power budget of transmitters is high enough; otherwise, FSA is a simple and effective solution.

To sum up, this paper has the following three major contributions:
\begin{itemize}\leftmargin=0in
\item [(1)] We are the first to compare packet-based and connection-based random access protocols in low-power IoT status update systems with AoI requirements. Specifically, for a given transmit power budget, we study different random access protocols to achieve high information freshness. 
\item [(2)] We use frame slotted Aloha (FSA) as the representatives of packet-based protocols and design request-then-access (RTA) as the connection-based protocol for theoretical analysis. Closed-form average AoI and average transmit power consumption formulas of different protocols are derived. Our study serves as a guideline for comparing packet-based and connection-based random access.
\item [(3)] We conduct comprehensive simulations to evaluate the performances of different protocols. The simulation results reveal that the favorability of connection-based random access  depends mainly on the payload size of the update data packets (compared with the request frames), as well as the transmit power budget. Overall, our investigations provide insights into the design of random access protocols for low-power timely status update systems. 
\end{itemize}

\section{Related Work}\label{sec:relatedwork}
\subsection{Age of Information (AoI)}
AoI was first proposed in vehicular networks to characterize the timeliness of safety packets \cite{aoi_first}. After that, it has been extensively studied under various communication and network systems; see the monograph \cite{AoI} and the references therein for important research results. Most early studies of AoI focused on the upper layers of the communication protocol stack, i.e., above the physical (PHY) and medium access control (MAC) layers. For example, a rich literature analyzed the AoI performance under different abstract queueing models, e.g., single-source single-server queues \cite{costa2016age}, multiple-source single-server queues \cite{yates2018age}, etc. To lower the network-wide AoI, age-optimal scheduling policies among multiple transmitters are examined in \cite{AoIscheduling1,AoIscheduling2,AoIscheduling3}, with the goal of minimizing different AoI metrics at a common receiver, such as average AoI and peak AoI \cite{AoIMagazine}. 

Moving down to the PHY and MAC layers, considering imperfect updating channels with interference and noise, the average AoI was analyzed in different network topologies, such as multi-hop \cite{aoi_multihop_ofdm} and multi-source \cite{multisource_error} networks. Moreover, different age-oriented error-correction techniques were investigated to combat the wireless impairments, e.g., automatic repeat request (ARQ) \cite{chenhe_aoi} and channel coding \cite{stream_code,arafa2019timely}. These works reveal that age-optimal designs are usually different from conventional delay/latency-optimal ones.

Energy is another crucial issue when designing AoI-aware status update systems, especially for low-power IoT sensor networks. For example, the AoI-energy characteristics of status update systems when using different ARQ protocols were discussed in \cite{chenhe_aoi,mangang_multicast,age_energy_truc_arq}. These works showed an inherent tradeoff between the AoI and the average energy consumption at the transmitters, especially under the generate-at-will model. In other words, the transmitter needs to decide whether to send (or resend) a packet under the transmit power constraint \cite{chenhe_aoi}. Unlike previous works, we do not consider ARQ in this paper. Instead, we study different MAC protocols in random access channels, with the goal of reducing the average AoI given an average transmit power budget. 

\subsection{Random Access Protocols related to AoI}
In the literature, different wireless access schemes on AoI were investigated, including scheduled access \cite{aoi_tdma_fdma,kuo2019minimum} and random access \cite{aoi_unrealible,aoi_irsa,adra}. Compared with scheduled access protocols, efficient random access protocols that operate in a distributed and decentralized way have shown promising results in IoT networks \cite{yang2021understanding,aoi_ultra_dense,aoi_random_access,aoi_crra}. Along this line, our work also focuses on AoI-aware random access protocols due to their importance and ease of implementation in distributed networks.

There has been rich literature on the modeling and performance analysis of AoI-aware random access protocols. In particular, slotted Aloha (SA)-based protocols have received the most attention. Despite being a relatively simple protocol, \cite{yang2021understanding} showed the AoI effectiveness of SA in massive access networks. Moreover, successive interference cancellation (SIC) was applied to improve the AoI performance of SA when the number of transmitters is large (i.e., the collision probability is high) \cite{aoi_irsa,grybosi2021age}. Moving beyond SA, the AoI of carrier sensing multiple access (CSMA) and general random access protocols were analyzed in \cite{aoi_csma} and \cite{aoi_ra}, respectively. Compared with the literature, this paper further investigates the average AoI of frame slotted Aloha (FSA), a generalization of SA. The AoI analysis of FSA was studied recently in \cite{FSA1, FSA2}. Specifically, \cite{FSA1} investigated the effects of retransmission on the average AoI of FSA, and \cite{FSA2} derived analytical expressions for the average and variance of AoI over a typical transmission link in Poisson bipolar and cellular networks. However, the above works focused extensively on the packet-based random access \cite{gao2019random}. 

To the best of our knowledge, the comparison between packet-based and connection-based random access protocols has not been investigated in the literature. This paper designs a request-then-access (RTA) protocol as the representative of connection-based random access. In the IEEE 802.11 standards, the payload size of data packets should exceed a threshold to activate the RTS/CTS mechanism so that the connection establishment facilitates higher network throughput \cite{dot11std13}. Unlike conventional 802.11 networks, our work considers AoI as the performance metric rather than network throughput. We notice that a recent worok \cite{aoi_reservation} studied the AoI performance of a reservation-based random access scheme that is similar to RTA. However, a fixed-duration access phase was considered in \cite{aoi_reservation}, which possibly leads to time wastage when most of the transmitters collide in the reservation phase (similar to the request phase in RTA). In contrast, RTA considers an access phase with variable duration depending on the number of transmitters that successfully contended in the request phase. We examine how the request phase in RTA reduces transmission failure time and transmit power wastage to achieve timely and low-powered state updates.

\section{Age of Information Preliminaries}\label{sec:preliminaries}
\subsection{Age of Information (AoI)}\label{sec:preliminaries1}
\begin{figure}
\centering
\includegraphics[width=0.25\textwidth]{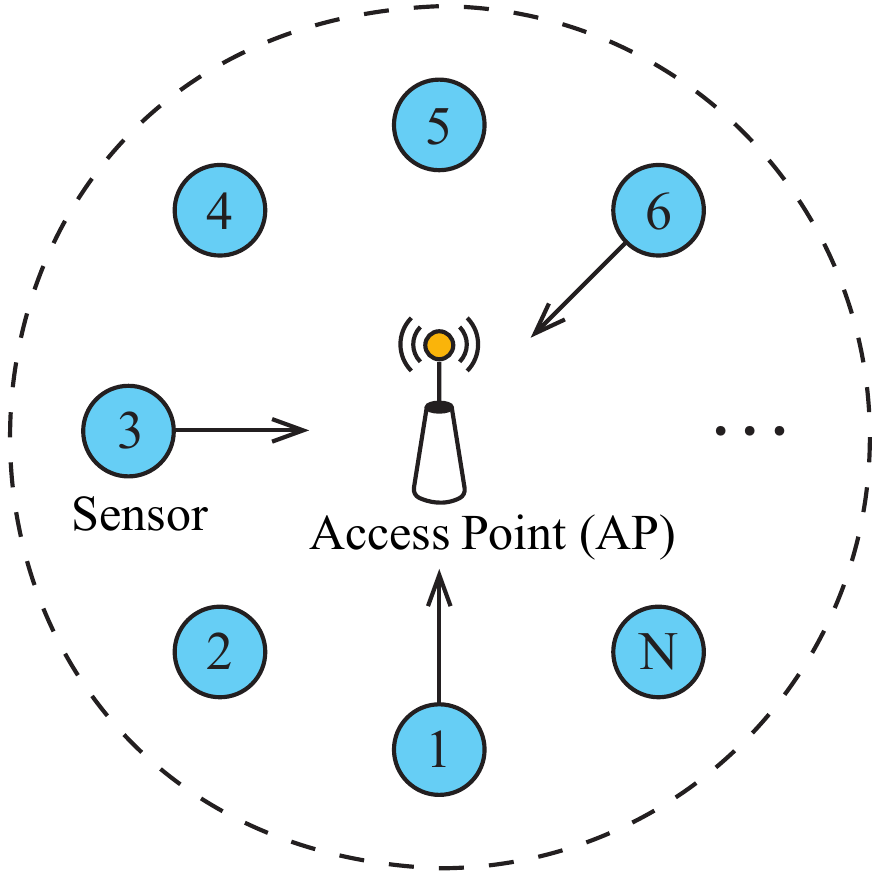}
\caption{A status update system with $N$ sensors sending update packets to the common access point (AP) in a random access manner.}
\label{fig:system_model}
\end{figure}

We study a timely status update system in which $N$ sensors send status update packets to a common access point (AP)\footnote{This paper assumes that the number of sensors, $N$, is known. In a practical random access network, it is difficult to obtain the exact $N$ at the receiver, which requires online estimation. The estimation of $N$ is out of the scope of the current work, and we refer interested readers to \cite{onlineSIC} for more details. The impact of an inaccurate $N$ on the AoI performance of different protocols will be presented in Section \ref{sec:performance}.}, as shown in Fig. \ref{fig:system_model}. 
At time instant $t$, the instantaneous AoI of sensor $u$, denoted by ${\Delta _u}(t)$, is defined by 
\begin{align}
{\Delta _u}(t) = t - {G_u}(t), 
\end{align}
where ${G_u}(t)$ is the generation time of the latest update packet received by the AP from sensor $u$ \cite{aoi_mono,AoI,yates2021age}. A lower ${\Delta _u}(t)$ means a higher degree of information freshness. 

With the instantaneous AoI ${\Delta _u}(t)$, we can compute the average AoI of sensor $u$. The average AoI ${\bar \Delta _u}$ is defined as the time average of the instantaneous AoI \cite{aoi_mono,AoI,yates2021age} 
\begin{align}
{\bar \Delta _u} = \mathop {\lim }\limits_{T \to \infty } \frac{1}{T}\int_0^T {{\Delta _u}(t)} dt.
\end{align}
A low average AoI ${\bar \Delta _u}$ indicates that the update packets from sensor $u$ are generally fresh over a long period. Considering the whole system, the average AoI for all the sensors is $\bar \Delta  = \frac{1}{N}\sum\nolimits_{u = 1}^N {{{\bar \Delta }_u}}$.

\begin{figure}
\centering
\includegraphics[width=0.4\textwidth]{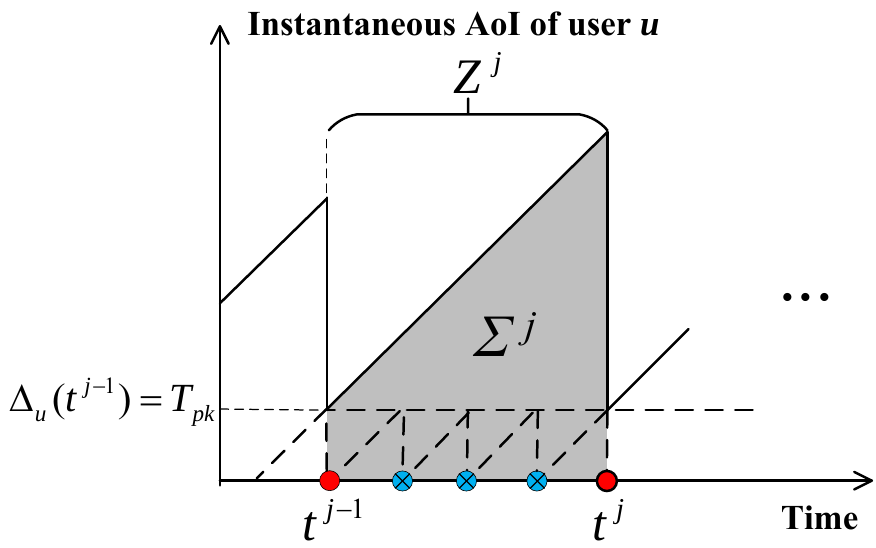}
\caption{An example of the instantaneous AoI ${\Delta _u}(t)$ of sensor $u$, where the $(j-1)$-th and the $j$-th updates occur at   ${t^{j - 1}}$ and ${t^{j}}$, respectively. We assume a generate-at-will model where a status packet is generated only when sensor $u$ has the opportunity to transmit. After a successful update, ${\Delta _u}(t)$ drops to ${T_{pk}}$ where ${T_{pk}}$ is the packet duration. }
\label{fig:general_ins_aoi}
\end{figure}

This paper considers a collision model, where a packet/frame is successfully received only when no other sensors' packets/frames are sent at the same time; otherwise, simultaneous transmissions from multiple sensors lead to a collision \cite{kurose2007computer}. Due to packet collision, sensor $u$ may send more than one update packet until the next successful update. Fig. \ref{fig:general_ins_aoi} plots an example of the instantaneous AoI ${\Delta _u}(t)$ in which the $(j-1)$-th and the $j$-th successful updates occur at ${t^{j - 1}}$ and ${t^{j}}$, respectively. We see that sensor $u$ sends four update packets after the last successful update at ${t^{j - 1}}$, and only the last one is successfully received by the AP at ${t^j}$. If an update packet is not received successfully, the instantaneous AoI ${\Delta _u}(t)$ continues to increase linearly. 

Let ${T_{pk}}$ denote the transmission time of an update packet. This paper considers the generate-at-will model, where a sensor can take measurements and generate a new update packet when it has the opportunity to transmit. The instantaneous AoI ${\Delta _u}(t)$ will drop to ${T_{pk}}$ when the AP successfully receives the update packet, e.g., at times ${t^{j - 1}}$ and ${t^{j}}$ as shown in Fig. \ref{fig:general_ins_aoi}. To realize the generate-at-will model in practice, the communication layer of the sensor can ``pull'' a request from the upper application layer just when there is an upcoming transmission opportunity (i.e., the packet generation and transmission is goal-oriented). This ensures that the sampled information is as fresh as possible, e.g., a sensor reading is obtained just before the transmission opportunity \cite{justintime}. 

Let us use $Z$ to represent the time between two consecutive status updates. To compute the average AoI ${\bar \Delta _u}$ of sensor $u$, let us consider area $\Sigma^j$ between the two consecutive successful updates in Fig. \ref{fig:general_ins_aoi}. The area of $\Sigma^j$ is calculated by
$\Sigma^j = {T_{pk}}{Z^j} + \frac{1}{2}{\left( {{Z^j}} \right)^2}$. 
Then the average AoI ${\bar \Delta _u}$ is computed by \cite{aoi_mono,AoI,yates2021age}
\begin{align}
{\bar \Delta _u} &= \mathop {\lim }\limits_{W \to \infty } \frac{{\sum\nolimits_{w = 1}^W {{\Sigma^w}} }}{{\sum\nolimits_{w = 1}^W {{Z^w}} }} = \frac{{\mathbb{E}\left[ {{T_{pk}}Z + \frac{1}{2}{{\left( Z \right)}^2}} \right]}}{{\mathbb{E}\left[ Z \right]}}\notag \\
&= {T_{pk}} + \frac{{\mathbb{E}\left[ {{Z^2}} \right]}}{{2\mathbb{E}\left[ Z \right]}}
\label{equ:avg_aoi_general}
\end{align}

\noindent where $\Sigma^{w}$ and $Z ^{w}$ denote the $w$-th $\Sigma$ and $Z$, respectively. In the following subsection, we use frame slotted Aloha (FSA) as an example to detail the computation of average AoI.

\subsection{The Average AoI of Frame Slotted Aloha (FSA)}\label{sec:FSA1}
\begin{figure}
\centering
\includegraphics[width=0.4\textwidth]{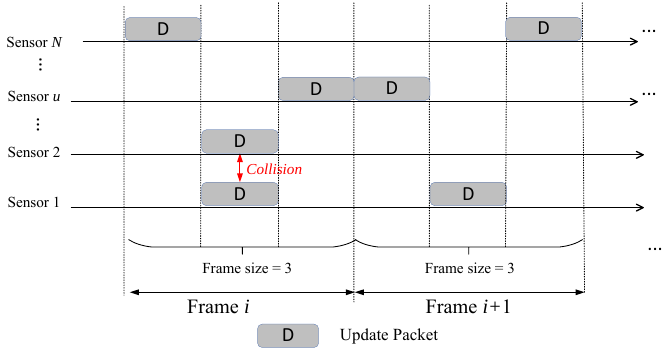}
\caption{An example of the FSA protocol. Each frame consists of three time slots. In every frame, each sensor sends an update packet with probability $\omega$ and randomly chooses one of the $k$ time slots to transmit.}
\label{fig:fsa_protocol}
\end{figure}

Time in FSA is divided into frames. Specifically, there are $k$ time slots in a frame for the sensors to contend and send update packets, e.g., $k=3$ time slots per frame in Fig. \ref{fig:fsa_protocol}. In each frame, a sensor sends an update packet with probability $\omega$; if a sensor has an update packet to send, it randomly chooses one of the $k$ time slots with equal probabilities (i.e., with probability $1/k$ for each time slot). In frame $i$ of Fig. \ref{fig:fsa_protocol}, four sensors send update packets to the AP. In particular, sensor 1 and sensor 2 select the same time slot, leading to a packet collision. The AP successfully receives the update packets from the remaining two sensors, sensor $u$ and sensor $N$. In frame $i+1$, three sensors send update packets to the AP without packet collision, i.e., all update packets are received by the AP.  In the rest of this paper, we also refer to a frame as a round. 

Since each sensor randomly chooses one of the $k$ time slots in a frame, let $D$ denote the time slot index that sensor $u$ chooses to send its update packet. Then $D$ has a probability mass function (PMF) 
\begin{align}
\Pr (D = d) = \frac{1}{k},~d = 1,2,...,k. 
\end{align}
In Fig. \ref{fig:fsa_protocol}, for example, we see $D = 1$ and $D = 3$  for sensor $N$ in frame $i$ and frame $i+1$, respectively. Let $D^{j - 1}$ and $D^{j}$ denote the time slot indices of the packets in the ($j-1$)-th update and the $j$-th update, respectively. We use $X$ to represent the number of rounds (frames) required for the next successful update. Thus, $Z$ is computed by 
\begin{align}
Z = (k{T_{pk}})X + (D^j - D^{j - 1}){T_{pk}}.
\end{align}

To compute the average AoI of FSA by (\ref{equ:avg_aoi_general}), we need to compute $\mathbb{E}[Z]$, which is
\begin{align}
\mathbb{E}[Z] = k{T_{pk}}\mathbb{E}[X] + \mathbb{E}[D^j - D^{j - 1}]{T_{pk}} = k{T_{pk}}\mathbb{E}[X],
\label{equ:fsa_exp_z}
\end{align}
\noindent where $D^{j - 1}$ and $D^{j}$  have the same distribution, i.e., $\mathbb{E}[D^j] = \mathbb{E}[D^{j - 1}] = \mathbb{E}[D]$. Similarly, the second moment of $Z$, $\mathbb{E}[{Z^2}]$, is computed by 
\begin{align}
\mathbb{E}[{Z^2}] &= \mathbb{E}\left[ {{{\left( {(k{T_{pk}})X + (D^j - D^{j - 1}){T_{pk}}} \right)}^2}} \right] \notag \\
&= {(k{T_{pk}})^2}\mathbb{E}[{X^2}] + \mathbb{E}\left[ {{{(D^j - D^{j - 1})}^2}} \right]{T_{pk}}^2\notag \\
& = {k^2}{T_{pk}}^2\mathbb{E}[{X^2}] + 2\left( {\mathbb{E}[{D^2}] - \mathbb{E}{{[D]}^2}} \right){T_{pk}}^2.
\label{equ:fsa_exp_z2}
\end{align}

\textbf{Computation of $\mathbb{E}[X]$ and $\mathbb{E}[{X^2}]$:} Let $\Omega$ and $\bar \Omega$ denote the events that sensor $u$ does and does not send a packet in a frame, respectively; also denote by events $S$  and $\bar S$ that sensor $u$ does and does not have a successful update in a frame, respectively.  Let ${\rm B}$ denote the number of the other $N-1$ sensors which send an update packet. Then the probability of event $S$ is
\begin{align}
\Pr (S) &= \sum\nolimits_{\beta  = 0}^{N - 1} {{\Pr (S|{\rm B} = \beta ,\Omega )  }\Pr ({\rm B} = \beta |\Omega )} \Pr (\Omega ) \notag \\
&= \omega \sum\nolimits_{\beta  = 0}^{N - 1} {{{\left( {1 - \frac{1}{k}} \right)}^\beta }{
{N - 1}\choose
\beta 
}{\omega ^\beta }{{\left( {1 - \omega } \right)}^{N - 1 - \beta }}}\notag  \\
& = \omega {\left( {1 - \frac{\omega }{k}} \right)^{N - 1}}, 
\end{align}
where $\Pr ({\rm B} = \beta |\Omega )={
{N - 1}\choose \beta }{\omega ^\beta }{{\left( {1 - \omega } \right)}^{N - 1 - \beta }}$. Since each frame (round) is independent, $X$ is a geometric random variable with a parameter $\Pr(S)=\omega {\left( {1 - \frac{\omega }{k}} \right)^{N - 1}}$. Thus, with the properties of geometric distributions, we have
\begin{align}
\mathbb{E}[X] = \frac{1}{\Pr(S)},~~
\mathbb{E}[X^2] = \frac{2-\Pr(S)}{\left (\Pr(S) \right )^2}.
\label{equ:x_exp}
\end{align}

\textbf{Computation of $\mathbb{E}[D]$ and $\mathbb{E}[{D^2}]$:} We next compute $\mathbb{E}[D]$ and $\mathbb{E}[{D^2}]$ for (\ref{equ:fsa_exp_z2}). It is easy to figure out that
\begin{align}
\mathbb{E}[D]&=\sum\limits_{d = 1}^k {d\Pr (D = d) = \frac{1}{k}} \sum\limits_{d = 1}^k {d = \frac{{k + 1}}{2}},\\
\mathbb{E}[D^2]&=\sum\limits_{d = 1}^k {{d^2}\Pr (D = d)}  = \frac{1}{k}\sum\limits_{d = 1}^k {{d^2} = \frac{{\left( {k + 1} \right)\left( {2k + 1} \right)}}{6}}.\notag
\end{align}
Now we can calculate the average AoI of FSA $\bar \Delta _u^{FSA}$ by (\ref{equ:aoi_fsa}).

We further consider the average transmit power consumption per user, which is defined as the ratio of the total energy cost to the total time consumed for the next successful update. Suppose that each sensor has a constant transmit power $P$. We can compute the average transmit power consumption of FSA, $\bar \Delta _u^{FSA}$, by (\ref{equ:aoi_fsa}), where $E_w$ is the total energy and $Z_w$ is the time consumed for the $w$-th successful update since the ($w$-1)-th successful update.

\newcounter{mytempeqncnt}
\begin{figure*}[!t]
\setcounter{mytempeqncnt}{\value{equation}}
\footnotesize
\begin{align}
\bar \Delta _u^{FSA}  = {T_{pk}} + \frac{{\mathbb{E}[{X^2}]{k^2}{T_{pk}}^2 + 2\left( {\mathbb{E}[{D^2}] - \mathbb{E}{{[D]}^2}} \right){T_{pk}}^2}}{{2k{T_{pk}}\mathbb{E}[X]}}
 = {T_{pk}} + k{T_{pk}}\frac{{2 - \omega {{\left( {1 - \frac{\omega }{k}} \right)}^{N - 1}}}}{{2\omega {{\left( {1 - \frac{\omega }{k}} \right)}^{N - 1}}}} + {T_{pk}}\frac{{\omega {{\left( {1 - \frac{\omega }{k}} \right)}^{N - 1}}\left( {{k^2} - 1} \right)}}{{12k}}.
 \label{equ:aoi_fsa}
\end{align}
\begin{align}
\bar P^{FSA} &= \mathop {\lim }\limits_{W \to \infty } \frac{{\sum\nolimits_{w = 1}^W {{E_w}} }}{{\sum\nolimits_{w = 1}^W {{Z_w}} }}= \frac{{\Pr (\Omega |\bar S){T_{pk}}(\mathbb{E}[X] - 1) + {T_{pk}}}}{{k{T_{pk}}\mathbb{E}[X]}}P 
 = \frac{{\frac{{\Pr (\bar S|\Omega )\Pr (\Omega )}}{{\Pr (\bar S)}}(\mathbb{E}[X] - 1) + 1}}{{k\mathbb{E}[X]}}P \notag\\
& = \frac{{\frac{{\left( {1 - {{\left( {1 - \frac{\omega }{k}} \right)}^{N - 1}}} \right)\omega }}{{1 - \omega  + \left( {1 - {{\left( {1 - \frac{\omega }{k}} \right)}^{N - 1}}} \right)\omega }}\left( {1 - \omega {{\left( {1 - \frac{\omega }{k}} \right)}^{N - 1}}} \right) + \omega {{\left( {1 - \frac{\omega }{k}} \right)}^{N - 1}}}}{k}P
 = \frac{wP}{k}.
\label{equ:fsa_average_power}
\end{align}
\hrulefill
\end{figure*}

In FSA, when more than one sensor chooses the same time slot, none of the sensors can successfully update and it is a waste of time to send the entire update packet, especially when the update packet duration is long. To address this issue, intuitively, we can let the sensors send a short control frame to contend for the channel (e.g., a RTS frame in the IEEE 802.11 standards \cite{dot11std13}) instead of sending a long data packet directly. Afterward, only successfully contended sensors can access the channel and send update packets. As the length of a request frame is typically smaller than that of a data packet, transmission failure time can be reduced. Motivated by this idea, we propose a request-then-access (RTA) random access protocol, described in the next section.

\section{Request-Then-Access Protocol}\label{sec:RTA}
This section details the request-then-access (RTA) protocol. Section \ref{sec:RTA1} first presents the protocol details, including the operation of the request phase and the access phase. Afterward, we analyze the average AoI of the RTA protocol in Section \ref{sec:RTA2}. In addition, the average transmit power consumption is analyzed. 
\begin{figure}
\centering
\includegraphics[width=0.48\textwidth]{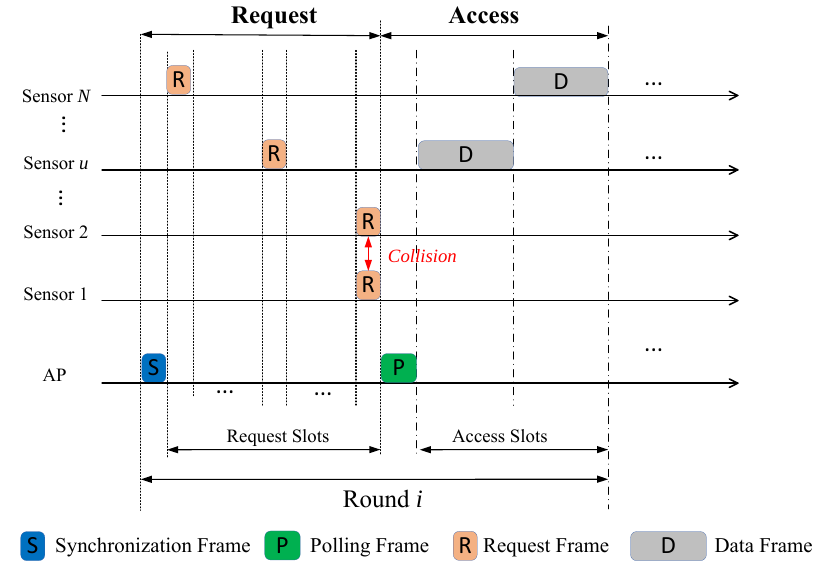}
\caption{The RTA protocol: in the request phase, a contention-based random access protocol is used by the sensors to send update requests to the AP; in the access phase, only the sensors that successfully contend in the request phase send update packets to the AP in a TDMA manner. }
\label{fig:rta_protocol}
\end{figure}

\subsection{Request-Then-Access (RTA) Protocol Description}\label{sec:RTA1}
Similar to FSA, time in the RTA protocol is divided into a series of rounds. While a round in FSA is a frame, a round in RTA consists of two phases, namely the request phase and the access phase, as shown in Fig. \ref{fig:rta_protocol}. We explain the two phases in detail below.

\textbf{\underline{Request Phase}}: This is a contention-based random access phase for the sensors to send update requests. In the request phase, the AP first broadcasts a synchronization frame at the beginning of a round to synchronize the time of different sensors. We omit the duration of the synchronization frame in the calculation of AoI because it is much shorter than the duration of a round. After that, there are $k$ request slots for the sensors to send a request frame. Specifically, in each round, a sensor transmits a request frame with probability $\pi$. As in FSA, if a sensor has a request frame to send, it randomly chooses one of the $k$ request slots with equal probabilities (i.e., probability $1/k$  for each request slot).
  
Fig. \ref{fig:rta_protocol} shows an example where four sensors send request frames in the request phase. Specifically, sensor 1 and sensor 2 choose the same request slot and their simultaneous request frames cause a collision. The AP successfully receives the request frames from the remaining two sensors (sensor $u$ and sensor $N$), so they are admitted to the subsequent access phase. We use ${T_r}$ to denote the duration of a request frame. In summary, the purpose of the request phase is to let the sensors contend for the channel. Successfully contended users establish a connection with the AP for sending data packets in the subsequent access phase.

\textbf{\underline{Access Phase}}: This is the update packet transmission phase for the sensors that contend successfully in the request phase. Initially, the AP sends a polling frame to notify the successful sensors to enter the access phase. The polling frame includes identifying the sensors and their order of sending update packets in the subsequent TDMA superframe. When calculating the AoI, we also omit the duration of the polling frame because it is short compared with a round.

For simplicity, we assume the order in which update packets are sent in the TDMA superframe is random. Let $M$ represent the number of sensors entering the access phase, and ${D_M}$ denote the time slot index of a particular sensor sending its update packet, given that $M$ sensors are allowed to enter the access phase. Then the PMF of ${D_M}$ is
\begin{align}
\Pr ({D_M} = d) = \frac{1}{M}, \  d = 1,2,...,M.
\end{align}

In Fig. \ref{fig:rta_protocol}, sensor $u$ and sensor $N$ are allowed to enter the access phase, so the access phase (i.e., the TDMA superframe) consists of two access slots. Each sensor has a probability of $1/2$ to send in the first access slot (i.e., $M=2$). In this example, sensor $u$ samples and sends a new update packet first, followed by sensor $N$. After both sensors send their update packets, the current round ends, and a new round.

\begin{figure}
\centering
\includegraphics[width=0.45\textwidth]{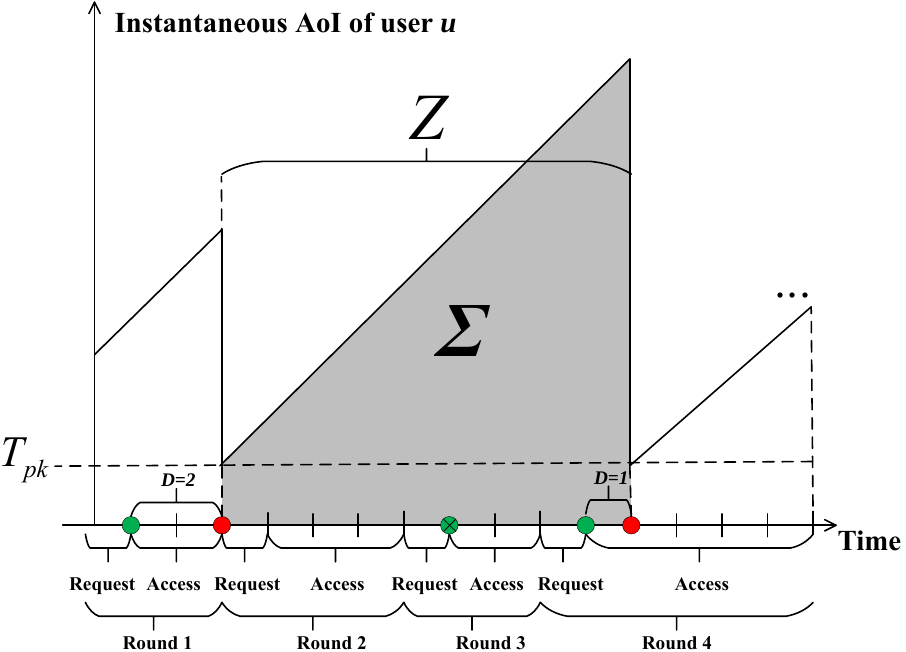}
\caption{An example of the instantaneous AoI in the RTA protocol. }
\label{fig:rta_ins_aoi}
\end{figure}

\subsection{Average AoI of the RTA Protocol}\label{sec:RTA2}
We now analyze the average AoI of the RTA protocol. Fig. \ref{fig:rta_ins_aoi} shows an example of the instantaneous AoI ${\Delta _u}(t)$ of sensor $u$ in RTA, where the first successful update occurs in round 1 and the second successful update occurs in round 4. Compared with FSA, the difficulties in analyzing the average AoI of RTA are as follows:
\begin{itemize}\leftmargin=0in
\item [(1)] The duration of a round in RTA is random. In particular, the duration of the access phase depends on the number of sensors $M$ in the access phase, i.e., the sensors that contend successfully in the request phase. As shown in Fig. \ref{fig:rta_ins_aoi}, $M=2$ in round 1 while $M=5$ in round 4 (in FSA, the duration of a round is always $k$ time slots).
\item [(2)] Since the number of sensors, $M$, in the access phase is random, ${D_M}$ is also random and depends on $M$. In Fig. \ref{fig:rta_ins_aoi}, we use the random variable $D$ as in FSA to denote the time slot index of the packet in the access phase, e.g., $D=2$ in round 1 while $D=1$ in round 4.
\item [(3)] After a successful update, suppose we use $X$ to denote the number of rounds spent until the next successful update. This means that sensor $u$ failed to update in the first $X-1$ rounds and then successfully updated in the last round. In Fig. \ref{fig:rta_ins_aoi}, $X=3$ after the first successful update. We need to calculate the duration of the $X-1$ rounds given that sensor $u$ failed to update (e.g., rounds 2 and 3 in Fig. \ref{fig:rta_ins_aoi}) and the duration of the last round given that sensor $u$ successfully updates (e.g., round 4 in Fig. \ref{fig:rta_ins_aoi}). Furthermore, notice that a failed update could be due to a collision of the request frames or a sensor simply remaining silent during a round.
\end{itemize}

As in FSA, let $Z$ denote the duration between two consecutive status updates, and $D^{j - 1}$ and $D^{j}$ denote the packet index of the ($j-1$)-th update and the $j$-th update in the TDMA superframe of the access phase, respectively. Then, $Z$ in RTA can be given by
\begin{align}
Z = \Theta _F^1 + ... + \Theta _F^{X - 1} + {\Theta _S} + (D^j - D^{j - 1}){T_{pk}},
\end{align}

\noindent where $\Theta _F^v,v = 1,...,X - 1,$ is the duration of a round given that sensor $u$ fails to have an update and ${\Theta _S}$ is the duration of a round given that sensor $u$ has a successful update. To compute the average AoI of RTA $\bar \Delta _u^{RTA}$ by (\ref{equ:avg_aoi_general}), $\mathbb{E}[Z]$ is computed by
 \begin{align}
\mathbb{E}[Z] &= (\mathbb{E}[X] - 1)\mathbb{E}[{\Theta _F}] + \mathbb{E}[{\Theta _S}] + \mathbb{E}(D^j - D^{j - 1}){T_{pk}}\notag \\
&= \left( {\mathbb{E}[X] - 1} \right)\mathbb{E}[{\Theta _F}] + \mathbb{E}[{\Theta _S}].
\label{equ:rta_z_exp}
\end{align}
\noindent The second equality is due to the fact that $\Theta _F^v,v = 1,...,X - 1,$ have the same distribution ($\mathbb{E}[\Theta _F^v] = \mathbb{E}[{\Theta _F}]$), and $\mathbb{E}[D^j] = \mathbb{E}[D^{j - 1}] = \mathbb{E}[D]$. 

Similarly, the second moment of $Z$,  $\mathbb{E}[{Z^2}]$, can be computed by (\ref{equ:rta_z_exp2}), where $\mathbb{E}[X]$ and $\mathbb{E}[X^2]$ are the same as (\ref{equ:x_exp}) except that is $\omega$ in FSA replaced by $\pi$ in RTA, i.e., the number of rounds until the next successful update $X$ is a geometric random variable with a successful update probability $\Pr (S) = \pi {\left( {1 - \frac{\pi }{k}} \right)^{N - 1}}$. We next compute the remaining components in (\ref{equ:rta_z_exp}) and (\ref{equ:rta_z_exp2}).

\begin{figure*}[!t]
\setcounter{mytempeqncnt}{\value{equation}}
\footnotesize
\begin{align}
\mathbb{E}[{Z^2}] &= \mathbb{E}\left[ {{{(\Theta _F^1 + ... + \Theta _F^{X - 1} + {\Theta _S} + (D^j - D^{j - 1}){T_{pk}})}^2}} \right]\notag \\
&= \mathbb{E}\left[ {{{\left( {\Theta _F^1 + ... + \Theta _F^{X - 1}} \right)}^2}} \right] + \mathbb{E}\left[ {{\Theta _S}^2} \right] + \mathbb{E}\left[ {{{(D^j - D^{j - 1})}^2}{T_{pk}}^2} \right] + 2\mathbb{E}\left[ {(\Theta _F^1 + ... + \Theta _F^{X - 1}){\Theta _S}} \right]\notag \\
&=\left( {\mathbb{E}[X] - 1} \right)\mathbb{E}[{\rm{ }}{\Theta _F}^2] + \left( {\mathbb{E}[{X^2}] - 3\mathbb{E}[X] + 2} \right)\mathbb{E}{[{\rm{ }}{\Theta _F}]^2} + \mathbb{E}[{\Theta _S}^2] + 2\left( {\mathbb{E}[X] - 1} \right)\mathbb{E}[{\Theta _F}]\mathbb{E}[{\Theta _S}] + 2\left( {\mathbb{E}[{D^2}] - \mathbb{E}{{[D]}^2}} \right){T_{pk}}^2
\label{equ:rta_z_exp2}
\end{align}
\hrulefill
\end{figure*}

\textbf{Computation of $\mathbb{E}[{\Theta _F}]$ and $\mathbb{E}[{\Theta _F}^2]$:} Since ${\Theta _F}$ is the duration of a round given that sensor $u$ fails to update, then ${\Theta _F} = k{T_r} + {M_F}{T_{pk}}$, where $M_F$ is the number of sensors admitted to the access phase given that sensor $u$ fails to have a successful update. The PMF of $M_F$ is
\begin{align}
\Pr ({M_F} = {m_F}) = &\Pr ({M_F} = {m_F}|\bar \Omega )\Pr(\bar \Omega ) \notag \\
&{\rm{ + }}\Pr ({M_F} = {m_F}|\Omega )\Pr(\Omega ).
\label{equ:prob_mf}
\end{align}
The first term of the right-hand side of (\ref{equ:prob_mf}) is the probability of $M_F$ given that sensor $u$ does not send a request frame (i.e., event $\bar \Omega$), and the second term is the probability of $M_F$ given that sensor $u$ does send a request frame (i.e., event $\Omega$). 

We first consider the case where sensor $u$ does not send a request frame. Let ${\rm A}$ denote the number of the remaining $N-1$ sensors which send a request frame. Given ${\rm A}= \alpha$, the number of sensors that contend successfully has a PMF
\begin{align}
&\Pr ({M_F} = {m_F}|{\rm A} = \alpha ,\bar \Omega ) \notag \\ 
&= \frac{{{{( - 1)}^{{m_F}}}k!\alpha !}}{{{k^\alpha }{m_F}!}} \times \sum\nolimits_{i = {m_F}}^{\min (k,\alpha )} {\frac{{{{( - 1)}^i}{{(k - i)}^{\alpha  - i}}}}{{(i - {m_F})!(k - i)!(\alpha  - i)!}}} ,\notag \\
&~~~~~\ 0 \le {m_F} \le \min (k,\alpha ).
\label{equ:mf_a_not_send}
\end{align}

Equation (\ref{equ:mf_a_not_send}) can be derived from the well-known combinatorial problem of assigning balls to boxes \cite{feller1991}. In that problem, a number of balls are thrown into boxes. Each box is selected with equal probability. Our problem is to assign ${\rm A} = \alpha$ sensors (balls) to $k$ request slots (boxes) and to calculate the probability that there are $m_F$ request slots with only one sensor 
in the $k$ request slots. For simplification in the rest of this paper, we use notation $\Pi ({m_F},\alpha ,k)$ to denote the right-hand side of (\ref{equ:mf_a_not_send}),
\begin{align}
&\Pi ({m_F},\alpha ,k) \notag= \\ &\frac{{{{( - 1)}^{{m_F}}}k!\alpha !}}{{{k^\alpha }{m_F}!}} \times \sum\nolimits_{i = {m_F}}^{\min (k,\alpha )} {\frac{{{{( - 1)}^i}{{(k - i)}^{\alpha  - i}}}}{{(i - {m_F})!(k - i)!(\alpha  - i)!}}} .
\end{align}
Considering all the possible ${\rm A}$, the number of sensors contended successfully given that sensor $u$ does not send a request frame has the following PMF
\begin{align}
\Pr ({M_F} = {m_F}|\bar \Omega ) &= \sum\nolimits_{\alpha  = {m_F}}^{N - 1} {\Pi ({m_F},\alpha ,k)} \Pr ({\rm A} = \alpha |\bar \Omega ),  \notag \\ 
&0 \le {m_F} \le \min (k,N - 1).
\end{align}

We next consider the case where sensor $u$ sends a request frame. Given $\alpha$ sensors sending requests, let ${\Phi _\alpha }$ denote the number of sensors (among the $\alpha$ sensors) that do not choose the same request slot as sensor $u$. The PMF of ${\Phi _\alpha }$ is
\begin{align}
\Pr ({\Phi _\alpha } = {\varphi _\alpha }|{\rm A} = \alpha ,\Omega ) = {
\alpha \choose
{{\varphi _\alpha }}
}{\left( {\frac{1}{k}} \right)^{\alpha  - {\varphi _\alpha }}}{\left( {1 - \frac{1}{k}} \right)^{{\varphi _\alpha }}}.
\end{align}
Given ${\Phi _\alpha } = {\varphi _\alpha }$, the number of successfully contended sensors in the case where sensor $u$ sends a request frame but that is corrupted has a PMF
\begin{align}
\Pr ({M_F} = {m_F}|{\Phi _\alpha } &= {\varphi _\alpha },{\rm A} = \alpha ,\Omega ) = \Pi ({m_F},{\varphi _\alpha },k - 1),\ \notag \\
&0 \le {m_F} \le \min (k - 1,{\varphi _\alpha }).
\label{equ:mf_a_send}
\end{align}

\begin{figure*}[!t]
\setcounter{mytempeqncnt}{\value{equation}}
\footnotesize
\begin{align}
\Pr ({M_F} &= {m_F}|\Omega ) = \sum_{\alpha  = {m_F}}^{N - 1} {\sum_{{\varphi _\alpha } = {m_F}}^\alpha  {\Pi ({m_F},{\varphi _\alpha },k - 1)} } \Pr ({\Phi _\alpha } = {\varphi _\alpha }|{\rm A} = \alpha ,\Omega )\Pr ({\rm A} = \alpha |\Omega )\notag \\
 &= \sum_{\alpha  = {m_F}}^{N - 1} {\sum_{{\varphi _\alpha } = {m_F}}^\alpha  {\Pi ({m_F},{\varphi _\alpha },k - 1)} } {
\alpha \choose
{{\varphi _\alpha }}
} {\left( {\frac{1}{k}} \right)^{\alpha  - {\varphi _\alpha }}}{\left( {1 - \frac{1}{k}} \right)^{{\varphi _\alpha }}} 
{{N - 1}\choose \alpha} 
{\pi ^\alpha }{\left( {1 - \pi } \right)^{N - 1 - \alpha }}, ~~~~0 \le {m_F} \le \min (k - 1,N - 1).
\label{equ:mf_send_final}
\end{align}
\hrulefill
\end{figure*}

Equation (\ref{equ:mf_a_send}) is to assign ${\Phi _\alpha } = {\varphi _\alpha }$ sensors to $k-1$ request slots (a request slot is occupied by user $u$ and $\alpha  - {\varphi _\alpha }$ sensors, which is a collision) and calculate the probability of having $m_F$ request slots with only one sensor in the $k-1$ request slots. Considering all the possible ${\Phi _\alpha }$ and ${\rm A}$, $\Pr ({M_F} = {m_F}|\Omega )$ can be computed by (\ref{equ:mf_send_final}). Therefore,  $\mathbb{E}[{\Theta _F}]$ and $\mathbb{E}[{\Theta _F}^2]$ are simply 
\begin{align}
\mathbb{E}[{\Theta _F}] &= k{T_r} + \mathbb{E}[{M_F}]{T_{pk}}, \\
\mathbb{E}[{\Theta _F}^2] &= {(k{T_r})^2} + {T_{pk}}^2\mathbb{E}[{M_F}^2] + 2k{T_r}{T_{pk}}\mathbb{E}[{M_F}].
\end{align}

\textbf{Computation of $\mathbb{E}[{\Theta _S}]$ and $\mathbb{E}[{\Theta _S}^2]$:} Since ${\Theta _S}$ is the duration of a round given that sensor $u$ has a successful update, then ${\Theta _S} = k{T_r} + {M_S}{T_{pk}}$, where ${M_S}$ is the number of sensors admitted to the access phase given that sensor $u$ has a successful update. Given ${\rm A} = \alpha$ of the remaining $N-1$ sensors sending request frames, the probability that all the $\alpha$  sensors do not choose the same request slot as sensor $u$ does is ${\eta _\alpha } = {\left( {1 - \frac{1}{k}} \right)^\alpha }$. Then the total number of sensors contend successfully given that sensor $u$ has a successful update has a PMF as (\ref{equ:ms_send_final}). 

\begin{figure*}[!t]
\setcounter{mytempeqncnt}{\value{equation}}
\footnotesize

\begin{align}
\Pr ({M_S} &= {m_S}) = \sum\nolimits_{\alpha  = {m_S} - 1}^{N - 1} {\Pi ({m_S} - 1,\alpha ,k - 1)} \Pr ({\rm A} = \alpha |S)\notag \\
= &\sum_{\alpha  = {m_S} - 1}^{N - 1} {\Pi ({m_S} - 1,\alpha ,k - 1){\pi{\left( {1 - \frac{1}{k}} \right)}^\alpha }} {
{N - 1}\choose
\alpha 
}{\pi ^{\alpha}}{\left( {1 - \pi } \right)^{N - 1 - \alpha }}\bigg/\left(\pi {\left( {1 - \frac{\pi }{k}} \right)^{N - 1}}\right), ~~~~1\le {m_S} \le 1 + \min (k - 1,N - 1).
\label{equ:ms_send_final}
\end{align}
\hrulefill
\end{figure*}

The meaning of (\ref{equ:ms_send_final}) is to assign ${\rm A} = \alpha$ sensors to $k-1$ request slots (a request slot is occupied by user $u$ only) and calculate the probability of having only one sensor in $m_S-1$ request slots among the remaining $k-1$ request slots, finally adding up all possible ${\rm A}$. Therefore, $\mathbb{E}[{\Theta _S}]$ and $\mathbb{E}[{\Theta _S}^2]$ are computed by 
\begin{align}
\mathbb{E}[{\Theta _S}] &= k{T_r} + \mathbb{E}[{M_S}]{T_{pk}},\notag \\
\mathbb{E}[{\Theta _S}^2] &= {(k{T_r})^2} + {T_{pk}}^2\mathbb{E}[{M_S}^2] + 2k{T_r}{T_{pk}}\mathbb{E}[{M_S}].
\end{align}
Next, we can compute the average AoI of RTA after calculating $\mathbb{E}[{D}]$ and $\mathbb{E}[D^2]$ with the PMF of $D$, i.e., 
\begin{align}
\Pr (D = d) = \sum\nolimits_{{m_S} = d}^{\min (k,N)} {\frac{1}{{{m_S}}}\Pr ({M_S} = {m_S})}.
\end{align}

Finally, we compute the average transmit power consumption of the RTA protocol $\bar P^{RTA}$. By definition,  $\bar P^{RTA}$ is computed by
\begin{align}
\bar P^{RTA} = \frac{{\Pr (\Omega |\bar S)(\mathbb{E}[X] - 1){T_r} + ({T_r} + {T_{pk}})}}{{(\mathbb{E}[X]-1)\mathbb{E}[{\Theta _F}]  + \mathbb{E}[{\Theta _S}]}}P,
\end{align}
\noindent where 
\begin{align}
\Pr (\Omega |\bar S) &= \frac{{\Pr (\bar S|\Omega )\Pr (\Omega )}}{{\Pr (\bar S)}} = \frac{{\left( {1 - {{\left( {1 - \frac{\pi }{k}} \right)}^{N - 1}}} \right)\pi  }}{{1 - \pi  + \left( {1 - {{\left( {1 - \frac{\pi }{k}} \right)}^{N - 1}}} \right)\pi  }} \notag \\
&=\frac{{\pi  - \pi {{\left( {1 - \frac{\pi }{k}} \right)}^{N - 1}}}}{{1 - \pi {{\left( {1 - \frac{\pi }{k}} \right)}^{N - 1}}}}
\end{align}
is the probability that sensor $u$ sends an update request given that the update is not successful due to collision. In other words, only when the sensor sends a request frame will consume power in the request phase. Since it is difficult to establish a clear relationship between $L^{RTA}$ and the average AoI $\bar \Delta _u^{RTA}$ as well as average power $\bar P^{RTA}$, we explore the performance of RTA by simulations as detailed in the next section.

\section{Performance Evaluation}\label{sec:performance}                            
In this section, we present numerical simulations to compare the performance of packet-based random access protocols (FSA) and the connection-based random access protocol (RTA). We conduct comprehensive simulations to evaluate the average AoI given an average transmit power consumption of different protocols. In particular, we will see how RTA improves the information freshness with a lower transmit power consumption.

\begin{table}[]
	\caption{\textcolor{black}{Network Simulation Parameters}}\label{table:notations_sim}
	\centering
	{\begin{tabular}{|  c|c|}
			\hline
			The number of sensors                         & 10, 15, 20, 25, 30, 40, 50, 60    \\ \hline
			Payload of an update packet  (bytes)                           &  8, 16, 24, 32, 40, 64, 128 \\ \hline
			Data transmission bitrate (Mbps)          &  6\\ \hline
			PHY-layer header duration ($\mu$s)    &  20\\ \hline
			MAC header + PHY pad (bits)    &  246\\ \hline
			Signal extension time ($\mu$s)    &  6\\ \hline
			Request frame (bits)    &  160\\ \hline
	\end{tabular}}
\end{table}

\begin{figure*}
\centering
\includegraphics[width=0.95\textwidth]{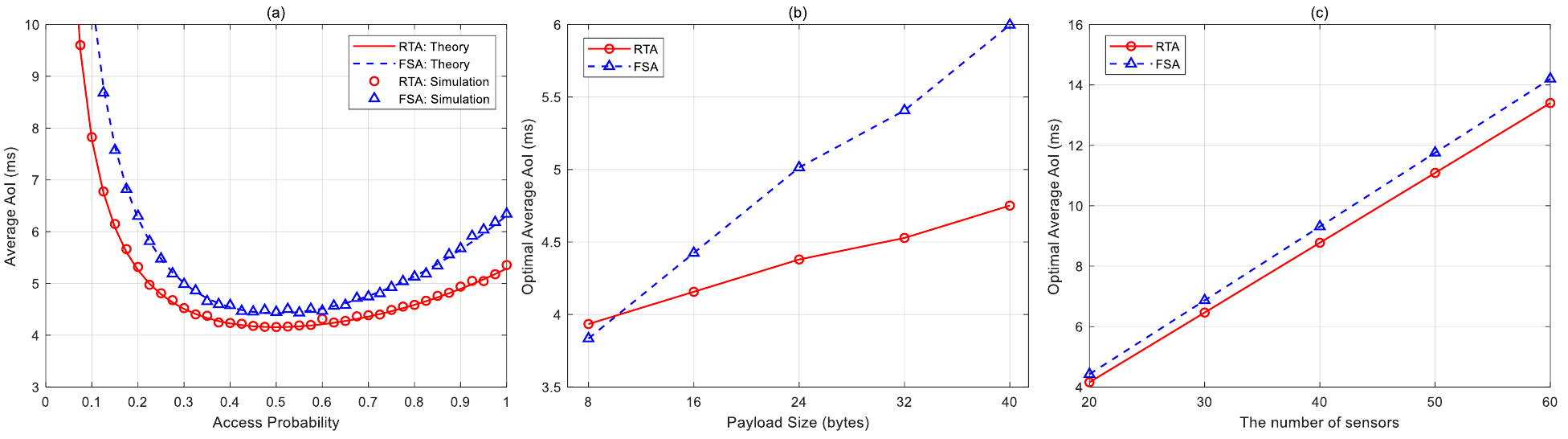}
\caption{Performance comparison between FSA and RTA: (a) the average AoI versus the access probability ($\omega$ in FSA and $\pi$ in RTA), (b) the optimal average AoI versus the payload of an update packet, and (c) the optimal average AoI versus the number of sensors. In (a) and (b), the number of sensors is 20. In (a) and (c), the payload of an update packet is 16 bytes. The numbers of time slots per frame in FSA and request slots in RTA are set to $k=10$.}
\label{fig:aoi_compare}
\end{figure*}

\subsection{Average AoI Comparison}\label{sec:performance1}
We first focus on the average AoI comparison. Table \ref{table:notations_sim} lists the network parameters of our simulations as defined in the IEEE 802.11 standards \cite{dot11std13}. To compute the average AoI of each protocol in our simulations, we first collect the instantaneous AoI of the sensors according to their successful updates or failed updates in each round over a long time. Next, we compute the average AoI based on the instantaneous AoI. 

Fig. \ref{fig:aoi_compare}(a) plots the average AoI of FSA and RTA versus the access probability ($\omega$ in FSA and $\pi$ in RTA), when an update packet has a 16-byte payload. In Fig. \ref{fig:aoi_compare}(a),  the number of sensors is fixed to $N=20$. The numbers of time slots per frame in FSA and request slots in RTA are set to $k=10$. We see that the simulation results corroborate the theoretical results for the average AoI, thus validating our average AoI derivations of FSA and RTA. 

As indicated in Fig. \ref{fig:aoi_compare}(a), the optimal average AoI of both FSA and RTA can be achieved when the access probability is not too low or too high. When the access probability is too low, the average AoI suffers due to the low frequency of sending update packets in FSA or requests in RTA (i.e., fewer update opportunities). When the access probability is too high, the average AoI is also subpar due to the high collision probability of the update packets in FSA or request frames in RTA (i.e., sensors cannot enter the access phase). The optimal access probability in Fig. \ref{fig:aoi_compare}(a) is 0.5. In addition, we can see from Fig. \ref{fig:aoi_compare}(a) that the optimal average AoI of RTA is lower than that of FSA. The primary reason for the performance improvement of RTA is that the duration of a request frame is smaller than that of an update packet. For example, when an update packet has a payload of 16 bytes, the duration of an update packet is around 90 $\mu s$ while the duration of a trigger frame is around 52$\mu s$. In this case, FSA wastes time when packets collide. In contrast, by establishing a connection first, RTA resolves collisions with a much shorter time in the request phase, and only the sensors which contend successfully send update packets later in the access phase. This improves channel utilization, which in turn improves the average AoI performance.

Fig. \ref{fig:aoi_compare}(b) investigates the effects of the payload size of an update packet on the optimal average AoI by plotting the optimal average AoI of FSA and RTA versus the payload size. We observe that FSA slightly outperforms RTA only when the payload size is as small as 8 bytes. When the payload size is larger than 16 bytes, the average AoI reduction of RTA compared to FSA increases with the payload size. Therefore, the use of FSA and RTA depends mainly on the payload size of data packets. Most of the time RTA outperforms FSA in average AoI because it helps avoid direct collision of data packets. The optimal average AoI versus the number of sensors is considered in Fig. \ref{fig:aoi_compare}(c). As shown in Fig. \ref{fig:aoi_compare}(c), the optimal average AoI of FSA and RTA increase with the number of sensors. Furthermore, Fig. \ref{fig:aoi_compare}(c) shows that the larger the number of sensors, the larger the average AoI improvement of RTA has over FSA, indicating that RTA is a viable solution to timely random access with massive sensors. 

\begin{figure}
\centering
\includegraphics[width=0.48\textwidth]{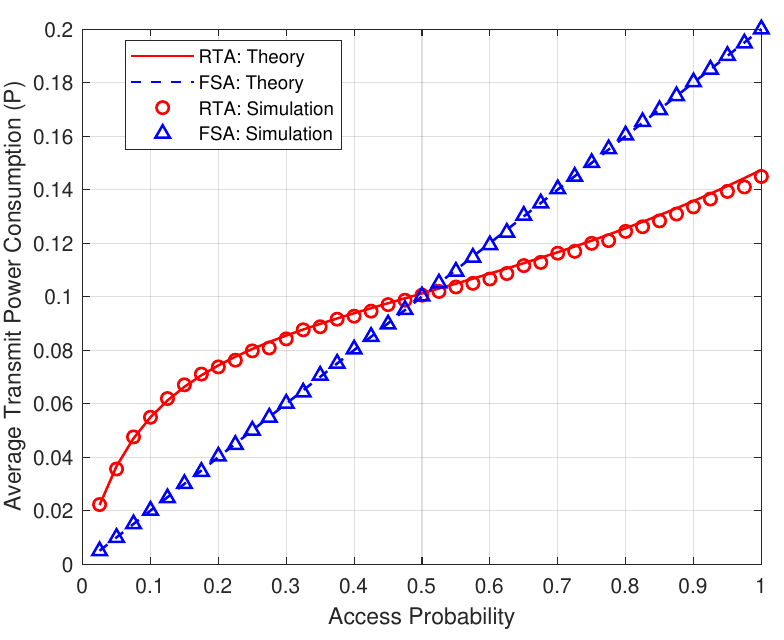}
\caption{Performance comparison between FSA and RTA: the average transmit power consumption versus the access probability. The number of sensors is 10. The payload of an update packet is 128 bytes. The numbers of time slots per frame in FSA and request slots in RTA are set to $k=5$.}
\label{fig: power}
\end{figure}

\begin{figure}
\centering
\includegraphics[width=0.48\textwidth]{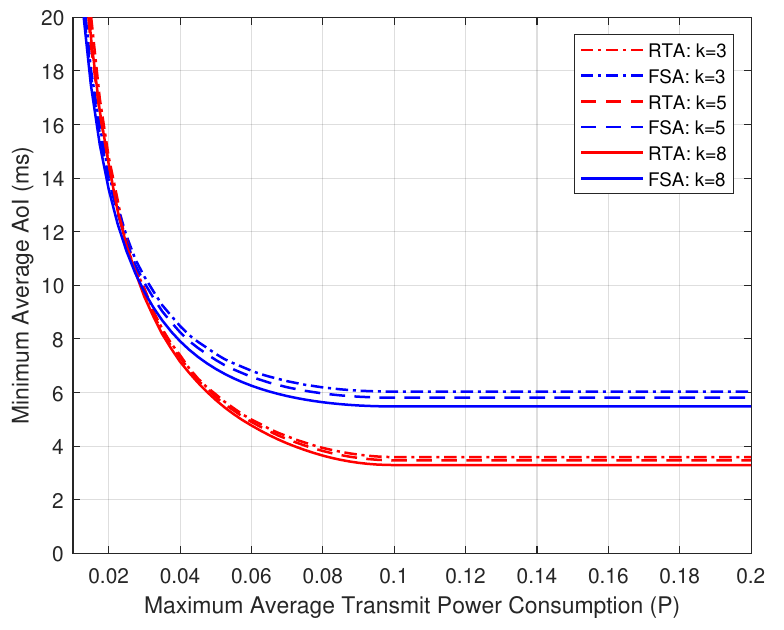}
\caption{Performance comparison between FSA and RTA: the minimum achievable average AoI versus the maximum average transmit power consumption. The number of sensors is 10. The payload of an update packet is 128 bytes. The figure plots different $k$ for RTA and FSA.}
\label{fig:aoi_power}
\end{figure}

\subsection{Minimum Average AoI Given A Maximum Average Transmit Power Consumption}\label{sec:performance2}
As mentioned earlier, the above simulation results do not take transmit power consumption into account. Fig. \ref{fig: power} plots the average transmit power consumption versus the access probability for different protocols ($\omega$ in FSA and $\pi$ in RTA). We consider a scenario where the number of sensors is 10 and the payload of an update packet is 128 bytes. The number of time slots per frame in FSA and the number of request slots in RTA is set to $k=5$. Simulation results verify the correctness of our theoretical analysis. Furthermore, we see from Fig. \ref{fig: power} that when the access probability is low, RTA has a higher average transmit power consumption than that of FSA. This is because when collision probability is low, RTA consumes additional power in sending request frames. On the contrary, when the access probability is large, the average transmit power consumption of RTA is lower than that of FSA. That is, when the collision probability is high, RTA consumes less power because sensors send only a short request frame instead of a long data packet directly in FSA.

We now evaluate the minimum average AoI of different protocols $\bar \Delta _{\min }(\bar P)$, given a maximum average transmit power $\bar P$. Note that we only provide theoretical results hereafter as their correctness has been verified. We first consider a scenario with the same setting as Fig. \ref{fig: power}, except that we also vary $k$. Fig. \ref{fig:aoi_power} plots $\bar \Delta _{\min }(\bar P)$ versus $\bar P$ of FSA, and RTA. We find that different $k$ have little effect on $\bar \Delta _{\min }(\bar P)$, so we simply choose $k=5$ for the rest of simulations, since the number of available time slots is typically fixed and smaller than the number of active users in random access scenarios. A fixed $k$ is also helpful for a time-varying number of active users and simplifies the system design in practice. As shown in Fig. \ref{fig:aoi_power}, for both protocols, as $\bar P$ increases, $\bar \Delta _{\min }(\bar P)$ first decreases and finally reduces to the minimum average AoI. We see that when the power budget is high, RTA has a substantial improvement on $\bar \Delta _{\min }(\bar P)$ over both FSA. For example, when $\bar P = 0.1P$, RTA reduces $\bar \Delta _{\min }(\bar P)$ by around 40\% compared with FSA. Hence, RTA is a viable solution to achieve higher information freshness when there is enough transmit power budget. 

\begin{figure}
\centering
\includegraphics[width=0.48\textwidth]{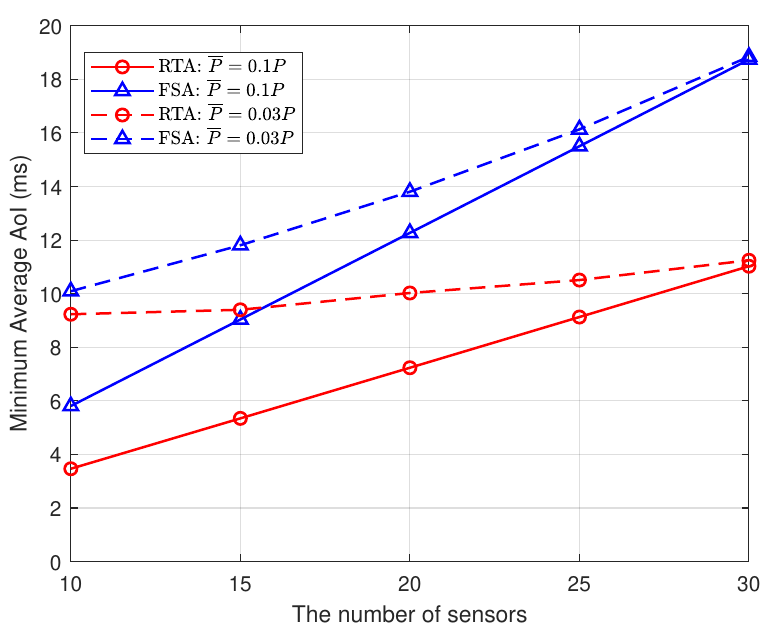}
\caption{Performance comparison between FSA and RTA: the minimum achievable average AoI versus the number of sensors, when the transmit power budget is $0.1P$ and $0.03P$. The payload of an update packet is 128 bytes.}
\label{fig:diff_users_128bytes}
\end{figure}

\emph{\underline{Impacts of the number of sensors}:} From Fig. \ref{fig:aoi_power}, we also see that RTA and FSA have almost the same $\bar \Delta _{\min }(\bar P)$ performance under a lower transmit power budget (e.g., when $\bar P = 0.03P$). To explore further, we plot $\bar \Delta _{\min }(\bar P)$ versus the number of sensors in Fig. \ref{fig:diff_users_128bytes}, when $\bar P$ equals $0.1P$ and $0.03P$, respectively. In both cases, the larger the number of sensors, the larger the AoI reduction of RTA compared with FSA. When the power budget is high,  $\bar \Delta _{\min }(\bar P)$ usually equals the minimum average AoI of the corresponding protocol. Hence, we see from Fig. \ref{fig:diff_users_128bytes} that RTA gives a lower average AoI than FSA does when $\bar P = 0.1P$. However, when the power budget is not high enough to achieve the minimum possible average AoI, FSA and RTA almost have the same average AoI when the number of sensors is small (e.g., 10 sensors when $\bar P = 0.03P$). In addition, Fig. \ref{fig:diff_users_128bytes} shows that RTA has a more stable average AoI than FSA under a different number of sensors. This indicates that even though $k$ is not optimized for both schemes, RTA can offer a more stable average AoI, thus significantly simplifying the system design.

\begin{figure}
\centering
\includegraphics[width=0.48\textwidth]{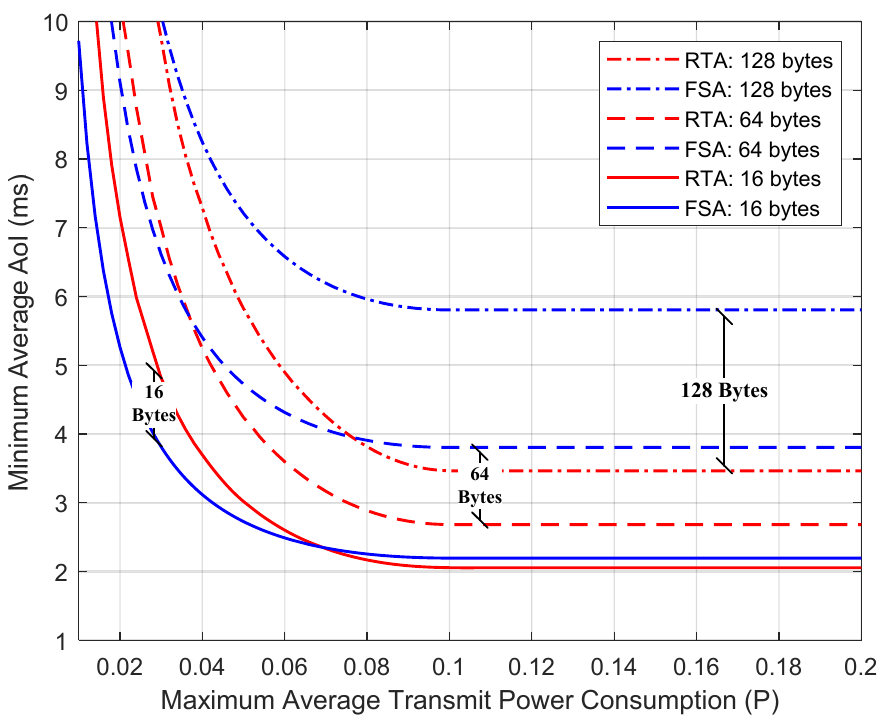}
\caption{Performance comparison between FSA and RTA: the minimum achievable average AoI versus the maximum average transmit power consumption under different payload sizes. The number of sensors is 10. The number of time slots per frame in FSA and the number of request slots in RTA are set to $k=5$.}
\label{fig:payload}
\end{figure}

\emph{\underline{Impacts of payload sizes}:} Fig. \ref{fig:payload} plots $\bar \Delta _{\min }(\bar P)$ versus $\bar P$ of FSA and RTA under different payload sizes, i.e., 16, 64, and 128 bytes, when the number of sensors is 10 (other numbers of sensors lead to similar phenomena, so we omit the results here). Previously in Fig. \ref{fig:aoi_power}, we show that when the payload of an update packet is 128 bytes and $\bar P = 0.1P$, RTA reduces $\bar \Delta _{\min }(\bar P)$ by 40\% compared with FSA. When the payload size is reduced to 64 bytes (the duration of an update packet is 156 $\mu s$), the performance improvement of RTA over FSA becomes smaller, i.e.,  RTA reduces $\bar \Delta _{\min }(\bar P)$ by 30\% compared with FSA. Furthermore, when the payload size is further reduced to 16 bytes, RTA still outperforms FSA by 6\%, when $\bar P = 0.1P$. When the payload size is 16 bytes, the duration of an update packet is only 90$\mu s$, which is comparable to the duration of a request frame (i.e., 52$\mu s$). Hence, when the probability of sending update requests is high, collisions in request slots (of RTA) do not differ much from collisions in regular time slots (of FSA). More specifically, when collisions occur, both waste almost the same amount of transmit power. 

Assuming that the given power budget $\bar P$ is as low as $0.03P$, we see that FSA achieves better performance than RTA when the payload of the update packet is 16 bytes, as shown in Fig. \ref{fig:payload}. Specifically, FSA reduces $\bar \Delta _{\min }(\bar P)$ by around 20\% compared with RTA. Intuitively, if the power budget is tiny, sensors should reduce the number of update requests per unit time to save power, i.e., reduce the probability of sending update requests. Since this reduces the total traffic and the packet collision probability, a dedicated request phase for collision resolution with $k$ request slots in RTA wastes the airtime, resulting in a higher average AoI than that of FSA. This indicates that under extremely low-power conditions, FSA is a preferred solution over RTA, especially when the update packet has a tiny payload.

To summarize, in low-power status update systems, whether a packet-based or connection-based random access protocol should be chosen mainly depends on the payload size of update packets and the transmit power budget of sensors. When the payload size is large (e.g., 128 bytes), a connection-based random access protocol such as RTA should be adopted because it dedicates a connection establishment phase to help avoid collisions of update packets, thus saving power and enhancing information freshness. When the payload size is small enough (e.g., 16 bytes) such that the duration of an update packet and a request frame are comparable, RTA should still be used if the power budget of sensors is high enough. Otherwise, a packet-based random access protocol such as FSA is a simple and effective solution, especially when the power budget is low.

\section{Conclusions}
We have compared the average AoI performance of packet-based and connection-based random access protocols under a maximum transmit power budget. Specifically, we use frame slotted Aloha (FSA) as an example of packet-based random access protocols. Motivated by FSA, we design a request-then-access (RTA) protocol for the analysis of connection-based random access protocols. 

The proper use of packet-based or connection-based random access protocols is of paramount importance in practical AoI-aware and low-power random access systems. Although prior works using throughput as the performance metric have shown that it is beneficial to establish connections when the payload size of data packets exceeds a threshold, how to determine an appropriate threshold from an AoI perspective has not been studied yet. Such an AoI characterization is much more complicated since AoI is generally related to the time elapsed between consecutive updates. Taking RTA as an example, the optimal probability of sending a request requires an in-depth analysis to minimize the average AoI. Besides, taking the transmit power consumption into account further complicates the problem. 

We have derived the closed-form average AoI and average transmit power consumption of FSA and RTA. Our theoretical analysis and simulation results indicate that connection establishment reduces the average AoI when the payload size is large. Specifically, RTA can reduce the average AoI by 40\% compared with FSA, when the update packet has 128 bytes. In contrast, FSA is a simple and effective solution when the power budget is low and the payload size is as small as 16 bytes. Our investigation on AoI can serve as a design reference for random access protocols with low-power and timely status update requirements. 


\bibliographystyle{IEEEtran}
\bibliography{low_power}

\vfill
\end{document}